\pdfoutput=1
\RequirePackage{latexml}
\iflatexml
	\documentclass{article}
	\author{
		Dominik Peters
		\\
		CNRS, LAMSADE, Universit\'e Paris Dauphine - PSL
		\\
		dominik.peters@lamsade.dauphine.fr
	}
\else
	\documentclass[11pt]{scrartcl}
	\usepackage[a4paper, total={16cm, 24cm}]{geometry}
	\renewcommand\tableofcontents{\listoftoc*{toc}} 
	
	\KOMAoptions{toc=bibnumbered}
	\setcapindent{0pt}
	\setkomafont{captionlabel}{\sffamily\bfseries}

	\usepackage{authblk}

	\author[1]{Dominik Peters}
	\affil[1]{CNRS, LAMSADE, Universit\'e Paris Dauphine - PSL}

	\usepackage{tocloft}

	\usepackage[small]{caption}
\fi

\usepackage{natbib}
\usepackage[pagebackref]{hyperref}
\hypersetup{
	pdfencoding=auto, 
	psdextra,
	colorlinks=true,
	citecolor=green!50!black,
	linkcolor=red!60!black,
	urlcolor=purple!70!black
}

\usepackage{url}
\newcommand{\showeprint}[2][]{%
	\href{https://arxiv.org/abs/#2}{arXiv:\nolinkurl{#2}}%
}
\usepackage{microtype}
\usepackage[T1,OT1]{fontenc}
\DeclareTextCommand{\k}{OT1}[1]{{%
	\fontencoding{T1}\selectfont\k{#1}%
}}
\usepackage[utf8]{inputenc}
\usepackage{graphicx}
\usepackage{amsmath}
\usepackage{amsthm}
\usepackage{amssymb}
\usepackage{mathtools}
\usepackage{booktabs}
\usepackage{enumitem}

\usepackage[nameinlink]{cleveref}
\usepackage{tikz}
\renewcommand{\epsilon}{\varepsilon}
\renewcommand*{\le}{\leqslant}

\renewcommand*{\ge}{\geqslant}
\renewcommand*{\geq}{\geqslant}

\newcommand{\github}{\raisebox{-1.5pt}{\begin{tikzpicture}
		\definecolor{black1}{HTML}{523d3f}
		\begin{scope}[scale=2.4]
			\fill[black1, even odd rule] (0.08,10)
			.. controls (0.03,10) and (0,9.97) .. (0,9.92)
			.. controls (0,9.89) and (0.02,9.86) .. (0.05,9.85)
			.. controls (0.06,9.85) and (0.06,9.85) .. (0.06,9.85)
			.. controls (0.06,9.85) and (0.06,9.86) .. (0.06,9.87)
			.. controls (0.04,9.86) and (0.03,9.87) .. (0.03,9.88)
			.. controls (0.03,9.88) and (0.03,9.88) .. (0.02,9.89)
			.. controls (0.02,9.89) and (0.02,9.89) .. (0.02,9.89)
			.. controls (0.03,9.89) and (0.03,9.89) .. (0.04,9.88)
			.. controls (0.04,9.87) and (0.05,9.87) .. (0.06,9.88)
			.. controls (0.06,9.88) and (0.06,9.89) .. (0.06,9.89)
			.. controls (0.05,9.89) and (0.03,9.9) .. (0.03,9.93)
			.. controls (0.03,9.93) and (0.03,9.94) .. (0.04,9.95)
			.. controls (0.04,9.95) and (0.03,9.96) .. (0.04,9.97)
			.. controls (0.04,9.97) and (0.04,9.97) .. (0.06,9.96)
			.. controls (0.07,9.96) and (0.07,9.96) .. (0.08,9.96)
			.. controls (0.08,9.96) and (0.09,9.96) .. (0.1,9.96)
			.. controls (0.11,9.97) and (0.12,9.97) .. (0.12,9.97)
			.. controls (0.12,9.96) and (0.12,9.95) .. (0.12,9.95)
			.. controls (0.13,9.94) and (0.13,9.93) .. (0.13,9.93)
			.. controls (0.13,9.9) and (0.11,9.89) .. (0.09,9.89)
			.. controls (0.1,9.88) and (0.1,9.88) .. (0.1,9.87)
			.. controls (0.1,9.86) and (0.1,9.85) .. (0.1,9.85)
			.. controls (0.1,9.85) and (0.1,9.85) .. (0.1,9.85)
			.. controls (0.13,9.86) and (0.16,9.89) .. (0.16,9.92)
			.. controls (0.16,9.97) and (0.12,10) .. (0.08,10)
			-- cycle;
		\end{scope}
\end{tikzpicture}}\,}

\newtheoremstyle{sfthm}
	{\topsep}
	{\topsep}
	{\itshape}
	{}
	{\sffamily\bfseries}
	{}
	{.5em}
	{}
\theoremstyle{sfthm}
\newtheorem{theorem}{Theorem}

\title{Kemeny Rank Aggregation \\ is NP-Hard for Three Voters}

\date{\vspace{-5pt}\sffamily\normalsize July 2026}

\newcommand{\KT}{\textsf{KT}}

\begin{document}

\maketitle

\begin{abstract}
	\iflatexml\else
	\begin{center}
		\textbf{\textsf{Abstract}} \smallskip
	\end{center}\fi
	Rank aggregation is the task of combining $n$ input rankings (linear orders) of alternatives into a single output ranking. The Kemeny rank aggregation rule selects the output ranking that minimizes the total Kendall-tau distance to the input rankings, i.e., the total number of adjacent swaps that need to be performed across input rankings so that they are all equal to the output ranking. \citet{dwork2001rank} proved that the problem of computing such a ranking is NP-complete for every even $n \ge 4$ and asked whether hardness holds even for $n = 3$. We give a hardness reduction from MAX CUT that proves the problem is NP-complete for $n = 3$.
	
	The reduction was found in July 2026 by GPT 5.6 Sol Ultra and simplified in part with help from Claude Fable 5.
\end{abstract}

\iflatexml\else
\vspace{15pt}

\hrule

\vspace{5pt}
{
	\setlength\columnsep{35pt}
	\setcounter{tocdepth}{2}
	\renewcommand\contentsname{\vspace{-20pt}}
		{\small
			\tableofcontents}
}

\vspace{11pt}
\hrule

\newpage
\fi

\section{Introduction}\label{sec:introduction}

Rank aggregation concerns the problem of combining several rankings into one ranking. If $A$ is a finite set of \emph{alternatives} (which we will also call \emph{candidates}), a \emph{ranking} $\succ$ is a linear order over $A$. Thus, the task is to find a map $(\succ_1, \dots, \succ_n) \mapsto {\succ}$ that takes as input a profile of $n$ rankings and returns a single ``summary'' or ``consensus'' ranking. Many applications have been discussed, including meta search engines, combining university rankings, and preference aggregation.

Given two rankings $\succ$, $\succ'$ over $A$, the \emph{Kendall-tau} distance between them is
\[
\KT(\succ, \succ') = |\{ (a,b) \in A \times A : a \succ b \text{ and } b \succ' a \}|,
\]
the number of pairs of alternatives on which the two rankings disagree. 
%
\citet{kemeny1959} introduced what is probably the most famous rank aggregation method,%
\footnote{This method is frequently studied thanks to its social choice properties \citep[see, e.g.,][]{Youn95a,younglevenglick,can2013update,bossert2014}. Its computation has also been studied from many perspectives, such as approximation algorithms \citep{coppersmith2010ordering,van2009deterministic,KMSc07a} and parameterized algorithms \citep{betzler2009}. Hardness is known even for 2-dimensional Euclidean preferences \citep{escoffier2022weighted}, and even for the problem of deciding whether a given output ranking is optimal \citep{FiHe22}.}
now known as the \emph{Kemeny method}. Given a profile $(\succ_1, \dots, \succ_n)$ it selects the ranking $\succ$ that minimizes 
\[
\text{Kemeny}_1(\succ) = \sum_{i \in [n]} \KT(\succ_i, \succ),
\]
i.e., the total Kendall-tau distance to the input rankings. This is a global optimization problem over $|A|!$ possible output rankings $\succ$, so it is not surprising that it is NP-hard \citep{bartholdi1989voting}. In an influential paper, \citet{dwork2001rank} proved that the problem is hard even if the number $n$ of input rankings is a constant, namely whenever $n \ge 4$ and $n$ is even. They noted that the ``complexity of computing a Kemeny optimal permutation for three full lists is open''. The problem of establishing Kemeny's complexity for $n = 3$ has remained open for 25 years. It has been called ``wide open'' \citep{biedl2009complexity}, ``quite intriguing'' \citep{fischer2016weighted}, ``famously open'' \citep{kraiczy2023weakly}, and ``long-standing'' \citep{jain2026bribery}.

To understand why extending the hardness result to $n = 3$ is challenging, it is useful to introduce a well-known equivalent formulation of the problem.
Given a profile, for any pair $a,b \in A$ of alternatives, write $n_{ab} = |\{ i \in [n] : a \succ_i b \}|$ for the number of voters who rank $a$ above $b$, and write $m_{ab} = n_{ab} - n_{ba}$ for the \emph{majority margin} of $a$ over $b$. If $m_{ab} > 0$ then a strict majority of voters prefer $a$ over $b$; if $m_{ab} = 0$ there is a tie; if $m_{ab} < 0$ then a strict majority of voters prefer $b$ over $a$. 

With this notation, we claim Kemeny selects the ranking $\succ$ minimizing the following quantity:
\[
\text{Kemeny}_2(\succ) = 
\textstyle
\sum_{(a,b) \in A \times A\, :\, a \succ b \text{ but } m_{ba} > 0} \: m_{ba}.
\]
In words, we look at all pairs of alternatives where $\succ$ disagrees with the majority (so it ranks $a \succ b$ but a majority of voters prefer $b$ to $a$) and imposes a cost of $m_{ba}$ for each such pair. The two objective functions $\text{Kemeny}_1(\succ)$ and $\text{Kemeny}_2(\succ)$ have the same minimizers, because one can check that
\[
\text{Kemeny}_1(\succ) = \sum_{\{a,b\} \subseteq A} \min(n_{ab}, n_{ba}) + \text{Kemeny}_2(\succ),
\]
and the first term is a constant that depends only on the profile, not the ranking $\succ$. Hence the two objective functions are positive affine transformations of each other, and thus have the same minimizers.

Now let us view the majority margins as an edge-weighted directed graph: for each pair $a,b$ of alternatives with $m_{ab} > 0$ we draw an arc from $a$ to $b$ with weight $m_{ab}$. (If $m_{ab} = 0$ we do not draw an arc.) Now, we can see that finding a ranking minimizing $\text{Kemeny}_2(\succ)$ is the same problem as finding a minimum-weight set of arcs that we have to reverse in order to make the graph acyclic. In other words, solving the Kemeny problem is just solving the (weighted) feedback arc set problem. Importantly, the reduction works in the other direction also: By the famous McGarvey--Debord theorem, every weighted (asymmetric) graph all of whose edge weights have the same parity can be induced as the margin graph of a ranking profile \citep{mcgarvey1953,debord1987} using a number of voters that is polynomial in the number of alternatives and the majority margins. In order to get a reduction from (unweighted) feedback arc set, we can just assign weight 2 to all the arcs of the input digraph, so all the weights even. Thus, because feedback arc set is NP-hard, so is Kemeny \citep{bartholdi1989voting}.

But notice that the margin graphs produced by the preceding hardness reduction have even margins, from which it follows that they can only be induced by ranking profiles with an even number of voters. Even numbers of voters allow for the possibility of a majority tie between alternatives ($m_{ab} = 0$) and thus allow for pairs of vertices that are not connected by an arc.

For an odd number of voters, the majority graph is always a weighted \emph{tournament}, i.e., there is an arc between \emph{every} pair of alternatives (directed one way or the other).
Making a hardness reduction work for feedback arc set on tournaments is quite challenging, and establishing NP-hardness for the ``feedback arc set on tournaments'' (FAST) problem was posed as an open problem by \citet{bang1992polynomial}. The problem remained open for 14 years, when \citet{alon2006tournaments} proved it NP-hard by derandomizing a randomized reduction for the FAST problem obtained by \citet[first published in 2005]{ailon2008aggregating}. Shortly thereafter, \citet{conitzer2006slater} discovered a delightful explicit reduction from the MAX SAT problem. It follows that computing Kemeny is hard even for an odd number of voters.

But how do we get from these reductions to showing that computing Kemeny is hard even for a \emph{constant} number of voters? What needs to be done is to show that the margin graphs appearing in the relevant hardness reductions can be induced by ranking profiles with a constant number of voters. For the even case, this is not too difficult, and was shown for $n = 4$ by \citet{dwork2001rank}, with \citet{biedl2009complexity} later correcting a minor error in the proof.%
\footnote{\label{fn:subdivision}\citet{bachmeier2019k} identified a simple reason why the $n=4$ reduction works: Take any graph and subdivide each edge. This does not change the size of the optimum feedback arc set. However, it can be shown that every subdivided graph can be induced using $n = 4$ voters.}
 (The case $n = 2$ is polynomial-time solvable. For example, both input rankings can be shown to be Kemeny rankings.) For the odd case, one can try to show that the tournaments appearing in \citeauthor{conitzer2006slater}'s \citeyearpar{conitzer2006slater} reduction can be induced by few voters. \citet{bachmeier2019k} did exactly this, showing that (a slight variant of) his reduction can be induced by $n = 7$ voters, providing the first constant odd number of voters for which Kemeny is hard. It is easy to see that hardness for a fixed $n$ also implies hardness for $n + 2$ (by adding two completely reversed orders to the profile, which does not change the optimum Kemeny ranking). Thus, we can deduce that Kemeny is hard for every even $n \ge 4$ and every odd $n \ge 7$.
 
 This left open the cases $n = 3$ and $n = 5$. A natural attempt to get hardness for these smaller $n$ is to further optimize the construction of \citet{bachmeier2019k}. But that seems difficult to do even for $n = 5$, and for $n = 3$ there are simple examples where the Conitzer tournaments just cannot be induced by three voters.%
 \footnote{For example, one can take the MAX SAT instance $\varphi=(x\vee y\vee z)\wedge(\neg x\vee y)\wedge(\neg y\vee z)\wedge(\neg z\vee x).$}
 Instead, in this paper, we establish hardness for $n = 3$ by constructing a new reduction showing that FAST is NP-hard. This reduction is designed exactly so that it \emph{can} be induced by $n = 3$ voters. The high-level shape of the reduction is still visibly similar to Conitzer's, but for us the interaction between variable gadgets and clause candidates is more complicated, and we use a different variable gadget that was designed using a SAT solver. (We will phrase our reduction as reducing from MAX CUT, so we will call them vertex gadgets and edge candidates.) Our result resolves the 25-year-old question of \citet{dwork2001rank} and shows that computing Kemeny is hard for every fixed $n \ge 3$.
 
 The reduction was found with the help of GPT 5.6 Sol Ultra (where ``Ultra'' specifies a mode available in OpenAI's Codex product where the model is encouraged to use many subagents) in July 2026. I had been asking LLMs for some time to prove hardness for the case of $n = 5$, with little success. GPT 5.6 Sol Ultra found a reduction for $n = 5$ in one shot after working autonomously for about 6 hours and 45 minutes. In my prompt, I had provided a copy of the paper by \citet{bachmeier2019k} as well as the paper by \citet{conitzer2006slater}. I also warned the model against trying to induce the same Conitzer reduction with $n = 5$ voters, since this is what models did in my prior prompting attempts and which is (in my view) a dead end. The reduction it found was pretty complicated, using a 26-candidate vertex gadget (Conitzer's reduction has 6 candidates in its gadget and so has the reduction presented in this paper) as well as additional ``wall'' gadgets for keeping the output ranking organized. I then went through many passes asking both Claude Fable 5 and GPT 5.6 to simplify the reduction. I then asked Claude Fable 5 to work towards the $n = 3$ case. It didn't make much progress, but left a variety of notes and scripts. Next, I asked GPT 5.6 Sol Ultra to look for a reduction with $n = 3$ voters, giving it the $n = 5$ reduction and Claude's notes. To my surprise, after roughly 4 hours of work, GPT reported that ``I think we may have the $n=3$ result'' (which was correct). After this breakthrough, there followed a lot of back-and-forth to try to find the simplest possible reduction, with me trying to explain to the LLMs (with limited success) what makes a reduction proof ``simple'' and ``nice''. My own main mathematical contributions lie in this simplification work (for example I suggested using the ``long even edges'' we will encounter below and which allow for a smaller vertex gadget and less case analysis). I also wrote this paper, since the current frontier models are really very bad at explaining their thinking to humans.
 
 Prompted by this work, I began building out a Lean library of formalized hardness reductions with the help of Claude models (\href{https://github.com/DominikPeters/HardnessReductionsLean/}{\github DominikPeters/HardnessReductionsLean}). We also formalized the correctness of the reduction presented in this paper
 (\href{https://github.com/DominikPeters/KemenyHardnessLean/}{\github DominikPeters/KemenyHardnessLean}).

\section{The Decision Problem}

Our reduction is from the following source problem, which is a restricted variant of SIMPLE MAX CUT (simple = unweighted edges). This problem gives us an undirected graph $\mathcal G = (\mathcal{V},\mathcal{E})$ and asks us to partition its vertices into a left-hand part $L$ and a right-hand part $R$ such that the number of edges that connect a left to a right vertex is as high as possible (such edges are called \emph{cut}). The non-simple version of MAX CUT was one of the original 21 NP-complete problems of \citet{karp1972}. The simple version was proved NP-complete by \citet{garey1976simplified}. We use a restricted version with ``long even edges''.
\medskip

\noindent
\begin{tabular}{p{0.98\linewidth}}
	\toprule
	SIMPLE MAX CUT with long even edges \\
	\midrule
	\textbf{Instance}: A graph $\mathcal G = (\mathcal{V},\mathcal{E})$ with $\mathcal{V} = \{1, \dots, n\}$, and a number $k \in \mathbb{N}$. The graph is simple (no repeated edges), undirected, and unweighted. Additionally, it satisfies the property that whenever $\{i,j\} \in \mathcal{E}$ then $|j - i| \ge 6$ and $i + j$ is even. \\
	\textbf{Question}: Does there exist a partition $L \cup R = [n]$ of the vertex set such that there are at least $k$ edges $e \in \mathcal{E}$ with $e \cap L \neq \emptyset$ and $e \cap R \neq \emptyset$? \\
	\bottomrule
\end{tabular}

\medskip
\noindent
Our version of this problem assumes ``long even edges'' to remove some case distinctions from the correctness argument. The problem is easily seen to be hard even with this restriction, by reduction from unrestricted SIMPLE MAX CUT: relabel each vertex $i$ as $6i$, relabel every edge $\{i,j\}$ as $\{6i, 6j\}$, and add additional isolated vertices to fill the resulting gaps in the vertex numbering. This operation does not change the optimal cut or the number of edges crossing it.

Our target problem can be stated as follows.
\medskip

\noindent
\begin{tabular}{p{15cm}}
	\toprule
	KEMENY SCORE with three voters \\
	\midrule
	\textbf{Instance}: A set $A$ of alternatives, three linear orders $\succ_1, \succ_2, \succ_3$ over $A$, and a number $b$. \\
	\textbf{Question}: Does there exist a linear order $\succ$ over $A$ such that $\text{Kemeny}_2(\succ) \le b$? \\
	\bottomrule
\end{tabular}

\medskip
\noindent
Our result is that this problem is NP-complete.

\begin{theorem}
	\textup{KEMENY SCORE with three voters} is NP-complete.
\end{theorem}

\noindent
Membership in NP is clear. Hardness follows from the following reduction.

\section{The Reduction}

Suppose we are given an instance of SIMPLE MAX CUT with long even edges: a graph $\mathcal G = (\mathcal{V},\mathcal{E})$ with $\mathcal{V} = \{1, \dots, n\}$, such that whenever $\{i,j\} \in \mathcal{E}$ with $i<j$, we have $j-i\geq 6$ and $i+j$ is even, and a number $k$.  We produce an instance of the KEMENY SCORE problem with three voters. Instead of immediately writing down the three linear orders in the input profile, we will begin by working with the majority graph, which will be a tournament on the candidate set, where each arc has weight 1. The cost $\text{Kemeny}_2(\succ)$ of an output ranking $\succ$ depends only on this tournament: it equals the number of pairs $a,b$ with an arc $a \to b$ but that are ranked $b \succ a$. Later, we will show that this tournament can be induced by a profile of 3 voters.

Set $M = 1000 (n+1)^4$. We think of $M$ as a large number, and of $M^2$ as a very large number.

We begin by specifying the candidate set. For each vertex $i \in \mathcal{V}$, we introduce six \emph{supercandidates} $A_i, B_i, C_i, D_i, E_i, F_i$, where each supercandidate is a set of $M$ fresh candidates. These six supercandidates will form a \emph{vertex gadget}
\[
	V_i = A_i \cup B_i \cup C_i \cup D_i \cup E_i \cup F_i.
\]
For each edge $e \in \mathcal{E}$, we introduce two candidates $\ell_e$ and $r_e$. The interpretation will be that $\ell_e$ ``wants'' one of the endpoints of $e$ to lie in the left part $L$, and $r_e$ wants one of the endpoints to lie in the right part $R$. Together, the candidate set is $\bigcup_{i=1}^n V_i \cup \{\ell_e, r_e : e \in \mathcal{E} \}$.

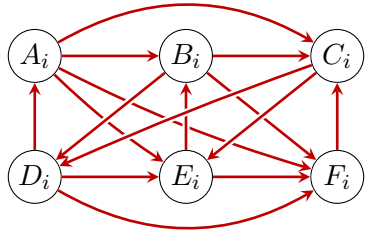
\begin{figure}[ht]
	\centering
	\begin{tikzpicture}[
		x=2.0cm,
		y=1.6cm,
		candidate/.style={circle,draw,fill=white,minimum size=7mm,inner sep=0pt},
		arc/.style={->,>=stealth,semithick},
		first cycles/.style={arc,draw=red!75!black,line width=1pt,
			preaction={draw=white,line width=2.5pt,-}},
		both cycles/.style={arc,draw=red!75!black,line width=1pt,
			preaction={draw=white,line width=2.7pt,-}},
		]
		\node[candidate] (A) at (0,1) {$A_i$};
		\node[candidate] (B) at (1,1) {$B_i$};
		\node[candidate] (C) at (2,1) {$C_i$};
		\node[candidate] (D) at (0,0) {$D_i$};
		\node[candidate] (E) at (1,0) {$E_i$};
		\node[candidate] (F) at (2,0) {$F_i$};
		
		\draw[first cycles] (A) to[bend right=5] (E);
		\draw[first cycles] (A) to[bend right=5] (F);
		\draw[first cycles] (B) -- (F);
		
		\draw[first cycles] (A) -- (B);
		\draw[first cycles] (B) -- (C);
		\draw[first cycles] (D) to[bend right=30] (F);
		
		\draw[first cycles] (A) to[bend left=30] (C);
		\draw[first cycles] (D) -- (E);
		\draw[first cycles] (E) -- (F);
		
		\draw[both cycles] (D) -- (A);
		\draw[both cycles] (E) -- (B);
		\draw[both cycles] (F) -- (C);
		\draw[both cycles] (B) -- (D);
		\draw[both cycles] (C) to[bend right=2] (D);
		\draw[both cycles] (C) -- (E);
	\end{tikzpicture}
	\caption{The tournament inside a vertex gadget.}
	\label{fig:vertex-gadget-tournament}
\end{figure}

Next, we specify the arcs of the majority tournament. 
Within each supercandidate $X_i$, we fix some arbitrary ordering of its $M$ members, say $X_i = \{ x_i^{(1)}, \dots, x_i^{(M)} \}$, and make a transitive tournament, so that $x_i^{(s)} \to x_i^{(t)}$ whenever $s < t$.
Between two supercandidates, or between a supercandidate and an edge candidate, all arcs will go in the same direction, so in the remainder of the construction, we can treat each supercandidate as if it is simply a candidate.
For a particular vertex $i \in \mathcal{V}$, its six supercandidates are arranged like in the tournament shown in \Cref{fig:vertex-gadget-tournament}: generally we have arcs
\[
A_i \to B_i \to C_i \to D_i \to E_i \to F_i
\]
and their transitive closure, \emph{except} that we reverse three of those arcs: 
\[
D_i \to A_i, \qquad E_i \to B_i, \qquad F_i \to C_i.
\]

For $i,j \in \mathcal{V}$, $i < j$, we put an arc from each supercandidate corresponding to $i$ to each supercandidate corresponding to $j$, i.e., $V_1 \to V_2 \to \dots \to V_n$ and its transitive closure.

We will not specify at this point the direction of the arcs between pairs of edge candidates; their direction does not matter for the correctness of the reduction. We will specify them later in a way that allows us to induce the tournament using three voters.

It remains to explain how edge candidates are connected to the vertex supercandidates. Fix an edge $e = \{i,j\}$ with $i < j$, and write $q=\frac{i+j}{2}$ for the midpoint. By the assumption that edges are even, $q$ is an integer. The arcs between $\ell_e$ and $r_e$ on the one side and a vertex gadget $V_u$ on the other are specified in \Cref{tbl:edge-candidate-beaten}. For example, it says that for $u = 1$ (assuming $i \neq 1$), candidate $\ell_e$ is beaten by ``all'' of $V_u$, which means $\{A_1, \dots, F_1\} \to \ell_e$, and it says that for $u = q$, candidate $\ell_e$ is beaten by ``$B_u,D_u,F_u$'', which means $\{B_q, D_q, F_q\} \to \ell_e \to \{A_q, C_q, E_q\}$.

\begin{table}[h]
	\centering
	$\begin{array}{l|ll}
			\toprule
			\text{gadget } V_u & \ell_e \text{ beaten by} & r_e \text{ beaten by}\\
			\midrule
			u = 1, \dots, i-1& \text{all} & \text{all}\\
			u = i & A_u&  A_u,B_u,E_u,F_u\\
			u = i+1, \dots,  q - 1& \text{none} & \text{none}\\
			u = q & B_u,D_u,F_u & B_u,D_u,F_u\\
			u = q + 1, \dots, j-1& \text{all} & \text{all}\\
			u = j &  A_u,B_u,C_u,D_u,E_u& C_u,D_u\\
			u = j+1, \dots, n & \text{none} & \text{none} \\
			\bottomrule
	\end{array}$
	\caption{Edge beating sets}
	\label{tbl:edge-candidate-beaten}
\end{table}

Finally, we set the target Kemeny cost to
\[
b = 3nM^2 + \sum_{\mathclap{\{i,j\}\in \mathcal{E},\ i<j}} \: \bigl(6\,(j-i)+3\bigr)M - kM + n^4.
\]

\section{Correctness}

\subsection{Forward Direction}

Suppose that $L \cup R$ is a partition of $\mathcal{V}$ with at least $k$ edges that are cut. We will assemble an output ranking $\succ$ with Kemeny cost at most $b$, which we will write as a string, so if we write $abc$ we mean $a \succ b \succ c$, and when $X$ and $Y$ are sets of alternatives, then if we write $XY$ we mean that $x \succ y$ for all $x \in X$ and all $y \in Y$. The output ranking will put the vertex gadgets in order $V_1V_2\dots V_n$. Within the vertex gadget $V_i$, we will place its supercandidates in the order
\begin{equation}
	A_iB_iC_iD_iE_iF_i \:\text{ if $i \in L$,} \qquad D_iA_iE_iB_iF_iC_i \:\text{ if $i \in R$.}
	\label{eq:optimal-gadget-orderings}
\end{equation}
Within each supercandidate, we rank its $M$ constituent candidates in their natural order.

Next, fix an edge $e=\{i,j\}$ with $i<j$.
Write $d = j-i$, which is even by assumption.
Place its two edge candidates according to \Cref{tab:edge-candidate-placements}, which is to be read as follows: The table contains four possible cases depending on whether $i$ is in $L$ or $R$ and whether $j$ is in $L$ or $R$. An entry
such as $A_i\,\ell_e\,B_i$ means that $\ell_e$ is placed between the two
supercandidates $A_i$ and $B_i$.  If several edge candidates are assigned
to the same gap, we order them arbitrarily.

\begin{table}[ht]
	\centering
	\small
	\setlength{\tabcolsep}{4pt}
	\begin{tabular}{cc@{\qquad}cc@{\qquad}ccc}
		\toprule
		\multicolumn{2}{c}{Vertex membership}
		& \multicolumn{2}{c}{Placement}
		& \multicolumn{3}{c}{Cost against vertex candidates} \\
		\cmidrule(lr){1-2}\cmidrule(lr){3-4}\cmidrule(lr){5-7}
		$i$ & $j$ & $\ell_e$ & $r_e$
		& $\ell_e$ & $r_e$ & Total \\
		\midrule
		$L$ & $L$
		& $A_i\,\ell_e\,B_i$ & $B_i\,r_e\,C_i$
		& $M(3d+2)$ & $M(3d+1)$ & $M(6d+3)$ \\
		$L$ & $R$
		& $A_i\,\ell_e\,B_i$ & $D_j\,r_e\,A_j$
		& $M(3d+2)$ & $M(3d)$ & $M(6d+2)$ \\
		$R$ & $L$
		& $E_j\,\ell_e\,F_j$ & $F_i\,r_e\,C_i$
		& $M(3d+2)$ & $M(3d)$ & $M(6d+2)$ \\
		$R$ & $R$
		& $A_i\,\ell_e\,E_i$ & $F_i\,r_e\,C_i$
		& $M(3d+3)$ & $M(3d)$ & $M(6d+3)$ \\
		\bottomrule
	\end{tabular}
	\caption{Placements and costs of the two edge candidates for
	$e=\{i,j\}$, according to the sides of its endpoints.}
	\label{tab:edge-candidate-placements}
\end{table}

Now we compute the Kemeny cost of this ranking. Within each supercandidate, there is no cost since the $M$ candidates in it are transitively ordered and the ranking respects all these internal arcs. Between the vertex gadgets of two vertices there is also no cost. Within each vertex gadget, there are exactly three pairs of supercandidates that are ranked incorrectly, namely
\[
\begin{cases}
	D_i \to A_i, \quad E_i \to B_i, \quad F_i \to C_i & \text{if $i \in L$}, \\
	B_i \to D_i, \quad C_i \to D_i, \quad C_i \to E_i & \text{if $i \in R$}.
\end{cases}
\]
Since each supercandidate consists of $M$ candidates, there are $M^2$ pairs of candidates that are misordered, incurring a total cost of $3M^2$ in each vertex gadget, or a total of $3nM^2$ across all vertex gadgets.

We haven't specified the arc directions between edge candidates, but in the worst case our ranking is wrong on all of them. In any case, this will incur a cost of at most $\binom{2|\mathcal{E}|}{2} \le n^4$.

Finally, we need to compute the cost incurred by edge candidates versus vertex candidates.
\Cref{tab:edge-candidate-placements} gives the relevant computations.
To understand where they are coming from, let's go through the first row (corresponding to the case $i,j\in L$) and let's compute the cost incurred by $\ell_e$ against vertex candidates. Let us scan through all vertices $u = 1, \dots, n$, and compute the cost of $\ell_e$ against candidates in $V_u$, by looking at who beats $\ell_e$ as specified in \Cref{tbl:edge-candidate-beaten}. 
\begin{itemize}
	\item There is no cost for $u = 1, \dots, i-1$ because those $V_u$ are placed \emph{before} $\ell_e$ in our ranking, which respects the fact that we have $V_u \to \ell_e$ in the tournament. 
	\item Inside $V_i$ there is also no cost because $A_i$ is placed before $\ell_e$ and the other five supercandidates are placed after it.
	\item The next $(\frac{d}{2} - 1)$ vertices (strictly between $i$ and the midpoint $\frac{i+j}{2}$) are also costless.
	\item At the midpoint, i.e. at $u = \frac{i+j}{2}$, $\ell_e$ is beaten by $3$ supercandidates, incurring cost $3M$.
	\item The next $(\frac{d}{2} - 1)$ vertices (strictly between the midpoint $\frac{i+j}{2}$ and $j$) have maximum cost because all $6$ supercandidates beat $\ell_e$ for a total cost of $6(\frac{d}{2} - 1)M = 3dM - 6M$.
	\item Finally, in $V_j$, there are $5$ supercandidates beating $\ell_e$, giving another $5M$ cost.
	\item The remaining vertices are again costless.
\end{itemize}
Summing up, we incurred cost $3M + 3dM - 6M + 5M = M(3d + 2)$, just as the table promised. The other cost entries can be checked similarly.
We can also use \Cref{tab:edge-candidate-placements} to confirm the intuition given earlier: the candidate $\ell_e$ wants one endpoint of $e$ to lie in $L$: whenever this is the case (the first three rows) it incurs a cost of only $M(3d+2)$ while otherwise it incurs a cost $M$ higher. Similarly, candidate $r_e$ wants one endpoint of $e$ to lie in $R$: whenever this is the case (the last three rows), it incurs a cost of only $M(3d)$ while otherwise it incurs a cost $M$ higher.

Overall, for each edge $e$, according to the table, the two candidates associated with $e$ have total cost $M(6d+2)$ against the vertex candidates when both edge candidates are happy (i.e., when $e$ is cut), and otherwise incur cost $M(6d+3)$.

Substituting back $d = j-i$, we can now compute an upper bound on the cost of our ranking:
\[
\underbrace{3nM^2}_{\clap{\parbox[c]{3cm}{\centering \scriptsize cost within \\ vertex gadgets}} }
\quad
+ 
\quad
\sum_{\mathclap{\{i,j\}\in \mathcal{E},\ i<j}} \: \bigl(6\,(j-i)+3\bigr)M - kM 
\quad + \quad
\underbrace{n^4}_{\clap{\parbox[c]{4cm}{\centering \scriptsize upper bound on cost \\ between edge candidates}}}
\]
which is exactly our cost budget $b$.

\subsection{Reverse Direction}
Now let's assume we are given a ranking $\succ$ with Kemeny cost at most $b$. Then the \emph{minimum} cost ranking also has Kemeny cost at most $b$, and so we may assume that the given ranking $\succ$ has minimum cost across all rankings. In addition, we may assume that for each supercandidate, its $M$ constituent candidates appear consecutively and in the right order inside the ranking $\succ$: that they appear consecutively follows from Theorem 1 of \citet{conitzer2006slater}, which shows that if a supercandidate is split into several blocks, we can move those blocks together in a way that doesn't worsen the Kemeny cost;%
\footnote{Here is a sketch of one proof of this result: view non-supercandidates as supercandidates of size 1, so the candidate set is partitioned into supercandidates. Given a ranking $\succ$, consider the following random process: for each supercandidate, select one of its members independently and uniformly at random, write $\succ'$ for the ranking obtained from $\succ$ by restricting to just the selected candidates, and then replace each candidate in $\succ'$ by a contiguous list of all the members of that candidate's supercandidate, obtaining ranking $\succ''$. One can show that the expected Kemeny cost of $\succ''$ equals the Kemeny cost of $\succ$. Hence there exists a ranking with all supercandidates present contiguously that has Kemeny cost at most that of $\succ$.} 
that they come in the right order follows because otherwise we could replace the whole block by the block ordered correctly and thereby reduce Kemeny cost.

Next, let's look into the cost incurred within a vertex gadget. Note that a vertex gadget contains three arc-disjoint 3-cycles:
\[
\underbrace{D_i \to A_i \to B_i \to D_i}_{\text{Cycle 1}}, \qquad
\underbrace{E_i \to B_i \to C_i \to E_i}_{\text{Cycle 2}}, \qquad
\underbrace{F_i \to C_i \to D_i \to F_i}_{\text{Cycle 3}}.
\]
For each of these cycles, the ranking $\succ$ must be wrong on at least one of its arcs; hence it is wrong on at least three arcs. Thus, the cost from within that vertex gadget alone is at least $3M^2$. In total across vertex gadgets, that gives within-gadget costs of at least $3nM^2$. Now suppose that $\succ$ got a fourth arc wrong in some vertex gadget. Then $\succ$ has a cost of at least $(3n + 1)M^2$. However, because $M$ is chosen very large, we have $(3n + 1)M^2 > b$, contradiction. Hence $\succ$ gets exactly three arcs wrong in each vertex gadget. It also must not interleave supercandidates from different vertex gadgets, and it must order the vertex gadgets in the right order $V_1 \succ \dots \succ V_n$, since otherwise it would again incur cost at least $(3n + 1)M^2 > b$.

We have established that $\succ$ gets exactly three arcs wrong within each vertex gadget. To see which orders can achieve this, also consider the following second collection of three arc-disjoint cycles:
\[
\underbrace{D_i \to A_i \to C_i \to D_i}_{\text{Cycle 4}}, \qquad
\underbrace{E_i \to B_i \to D_i \to E_i}_{\text{Cycle 5}}, \qquad
\underbrace{F_i \to C_i \to E_i \to F_i}_{\text{Cycle 6}}.
\]
\begin{figure}[t]
\centering
\begin{tikzpicture}[
    cycle/.style={rectangle, draw, rounded corners, fill=white, inner sep=3pt},
    incidence/.style={semithick}
]
    \node[cycle] (c1) at (-4,1.5) {Cycle 1};
    \node[cycle] (p1) at (0,1.52) {Cycle 4};
    \node[cycle] (c3) at (4,1.5) {Cycle 3};
    \node[cycle] (p3) at (4,-0.38) {Cycle 6};
    \node[cycle] (c2) at (0,-0.38) {Cycle 2};
    \node[cycle] (p2) at (-4,-0.38) {Cycle 5};

    \draw[incidence] (c1) -- node[sloped, fill=white, inner sep=1pt, text=red!73!black] {$D_i \to A_i$} (p1);
    \draw[incidence] (p1) -- node[ sloped, fill=white, inner sep=1pt, text=blue] {$C_i \to D_i$} (c3);
    \draw[incidence] (c3) -- node[fill=white, inner sep=1pt, text=red!73!black] {$F_i \to C_i$} (p3);
    \draw[incidence] (p3) -- node[ sloped, fill=white, inner sep=1pt, text=blue] {$C_i \to E_i$} (c2);
    \draw[incidence] (c2) -- node[sloped, fill=white, inner sep=1pt, text=red!73!black] {$E_i \to B_i$} (p2);
    \draw[incidence] (p2) -- node[fill=white, inner sep=1pt, text=blue] {$B_i \to D_i$} (c1);
\end{tikzpicture}
\caption{The cycle--arc incidence graph of the two collections of cycles. An edge represents an arc common to its incident cycles.}
\label{fig:cycle-arc-incidence}
\end{figure}
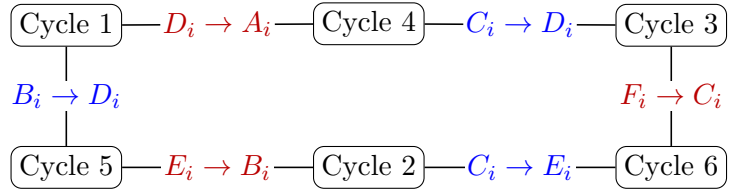
An order with only three wrong arcs must choose one arc from each cycle in both collections. Every one of its wrong arcs must therefore occur in one cycle of each collection. There are six arcs that appear in both collections. In \Cref{fig:cycle-arc-incidence}, we draw a graph where each cycle is a vertex and each of those six arcs is an edge connecting the two cycles in which it appears. We must select 3 arcs that hit all cycles; since the graph is a 6-cycle, there are exactly two possible sets of three wrong arcs that do so, namely the two perfect matchings of the graph (indicated in blue and red in the figure):
\[
\{D_i\to A_i,E_i\to B_i,F_i\to C_i\}
\quad\text{or}\quad
\{B_i\to D_i,C_i\to D_i,C_i\to E_i\}.
\]
After reversing either set of arcs, the resulting graph is acyclic and encodes a unique linear order. This process can yield either of the two orders in \eqref{eq:optimal-gadget-orderings}.%
\footnote{Instead of this fancy argument, one can also just enumerate all $6!=720$ orderings of the six supercandidates and compute the Kemeny costs of each. This enumeration will find exactly two optima: ABCDEF and DAEBFC.}
Hence, we can define
\begin{align*}
	L &= \{ i \in \mathcal V : A_i \succ B_i \succ C_i \succ D_i \succ E_i \succ F_i \}, \\
	R &= \{ i \in \mathcal V : D_i \succ A_i \succ E_i \succ B_i \succ F_i \succ C_i \},
\end{align*}
and these two sets form a partition of the vertex set of $\mathcal G$. 

It remains to argue that at least $k$ edges cross the cut $L$--$R$.
For this, we need to understand the placement of the edge candidates relative to the vertex gadgets, recalling that we have a complete understanding of how the vertex gadgets themselves are ordered.
We begin by figuring out the best location to insert the edge candidates -- the location that induces minimum cost between the edge candidate and all the vertex gadget candidates.
For this, we will scan through the vertex gadgets from left to right, and compute the costs at various positions using the beating sets as specified in \Cref{tbl:edge-candidate-beaten}. (In the following arguments we are going to neglect the relationships between different edge candidates since they make a negligible contribution to the score. Getting one more relationship between a vertex supercandidate and an edge candidate wrong, compared to the optimum, costs $M$ points, while the total impact of the relationships between edge candidates is bounded by $n^4 < M$.)

Fix some edge $e = \{i,j\} \in \mathcal E$ with $i < j$. Set $d=j-i$ and $q = \frac{i+j}{2}$.
First, the globally optimal location must also be a locally optimal location (the cost does not go down by moving the edge candidate one slot left or right).
It follows immediately that the optimum locations for $\ell_e$ and $r_e$ can only be somewhere within $V_i$ or $V_q$ or $V_j$: If such a candidate were placed in an intermediate vertex gadget, we would not have local optimality. For example, if placed inside $V_u$ for $u < i$, then since $V_u \to \{\ell_e, r_e\}$, the Kemeny cost would be reduced by pushing the edge candidate to the right towards $V_i$; similarly, for $i < u < q$, we can reduce by pushing the edge candidate to the left towards $V_i$. There is a similar attractor towards $V_j$: for $q < u < j$ we can reduce cost by pushing the edge candidate to the right towards $V_j$, and for $j < u$ we can reduce cost by pushing to the left towards $V_j$.

We next rule out that an edge candidate is placed within the midpoint gadget $V_q$. By assumption, $j-i\geq 6$, so there are at least two complete vertex gadgets strictly between $V_i$ and $V_q$. By \Cref{tbl:edge-candidate-beaten}, both edge candidates beat every supercandidate in those gadgets. Placing an edge candidate as far right as $V_q$ therefore incurs at least $12M$ more on these two gadgets than placing it immediately after $V_i$. Its position relative to the six supercandidates of $V_q$ affects at most $6M$ comparisons. Thus, placement in $V_q$ is strictly worse (by at least $12M-6M = 6M$) than placing it immediately after $V_i$; hence the optimum ranking does not place the edge candidate in $V_q$.

It follows that in the minimum-cost ranking, $\ell_e$ and $r_e$ are placed within $V_i$ or $V_j$. Once placed inside one of these gadgets, let us compute the total cost incurred between the edge candidate and all the \emph{other} vertex gadgets. For $\ell_e$, if placed in $V_i$, it incurs $3M$ cost against the midpoint gadget $V_q$, $6(\frac{d}{2}-1)M$ cost against the gadgets between the midpoint and $V_j$, and $5M$ cost against $V_j$, for a total of $(3d+2)M$. If placed in $V_j$, it incurs costs $3M$ against $V_q$, cost $6(\frac{d}{2}-1)M$ against the gadgets between $V_q$ and $V_i$, and cost $5M$ against $V_i$, which sums to the same amount: $(3d+2)M$. Similarly, one can compute that $r_e$ incurs a cost of $(3d-1)M$ against vertex gadgets other than the one it is placed inside, no matter whether it is placed inside $V_i$ or $V_j$.

\begin{figure}[ht]
	\centering
	\begin{tikzpicture}[
		x=.8cm,
		y=.8cm,
		supercandidate/.style={
			draw,
			rounded corners=1pt,
			fill=white,
			minimum width=.62cm,
			minimum height=.52cm,
			inner sep=1pt,
			font=\small
		},
		insertion/.style={->,>=stealth,semithick},
		ell insertion/.style={insertion,draw=blue!70!black},
		r insertion/.style={insertion,draw=red!70!black},
		ell cost/.style={font=\scriptsize,text=blue!70!black},
		r cost/.style={font=\scriptsize,text=red!70!black},
		normal/.style={},
		best/.style={font=\scriptsize\bfseries,draw=black,circle,fill=white,inner sep=.6pt}
	]
		\begin{scope}
			\node[font=\small\bfseries] at (3,1.55) {inside $V_i$ if $i\in L$};
			\foreach \x/\name in {.5/A,1.5/B,2.5/C,3.5/D,4.5/E,5.5/F}
				\node[supercandidate] at (\x,0) {$\name$};
			\node[ell cost] at (-.65,.92) {$\ell_e$};
			\node[r cost] at (-.65,-.92) {$r_e$};
			\foreach \x/\value/\style in {0/1/normal,1/0/best,2/1/normal,3/2/normal,4/3/normal,5/4/normal,6/5/normal} {
				\draw[ell insertion] (\x,.67) -- (\x,.30);
				\node[ell cost,\style] at (\x,.92) {$\value$};
			}
			\foreach \x/\value/\style in {0/4/normal,1/3/normal,2/2/best,3/3/normal,4/4/normal,5/3/normal,6/2/best} {
				\draw[r insertion] (\x,-.67) -- (\x,-.30);
				\node[r cost,\style] at (\x,-.92) {$\value$};
			}
		\end{scope}

		\begin{scope}[xshift=7.4cm]
			\node[font=\small\bfseries] at (3,1.55) {inside $V_i$ if $i\in R$};
			\foreach \x/\name in {.5/D,1.5/A,2.5/E,3.5/B,4.5/F,5.5/C}
				\node[supercandidate] at (\x,0) {$\name$};
			\node[ell cost] at (-.65,.92) {$\ell_e$};
			\node[r cost] at (-.65,-.92) {$r_e$};
			\foreach \x/\value/\style in {0/1/best,1/2/normal,2/1/best,3/2/normal,4/3/normal,5/4/normal,6/5/normal} {
				\draw[ell insertion] (\x,.67) -- (\x,.30);
				\node[ell cost,\style] at (\x,.92) {$\value$};
			}
			\foreach \x/\value/\style in {0/4/normal,1/5/normal,2/4/normal,3/3/normal,4/2/normal,5/1/best,6/2/normal} {
				\draw[r insertion] (\x,-.67) -- (\x,-.30);
				\node[r cost,\style] at (\x,-.92) {$\value$};
			}
		\end{scope}

		\begin{scope}[yshift=-3.5cm]
			\node[font=\small\bfseries] at (3,1.55) {inside $V_j$ if $j\in L$};
			\foreach \x/\name in {.5/A,1.5/B,2.5/C,3.5/D,4.5/E,5.5/F}
				\node[supercandidate] at (\x,0) {$\name$};
			\node[ell cost] at (-.65,.92) {$\ell_e$};
			\node[r cost] at (-.65,-.92) {$r_e$};
			\foreach \x/\value/\style in {0/5/normal,1/4/normal,2/3/normal,3/2/normal,4/1/normal,5/0/best,6/1/normal} {
				\draw[ell insertion] (\x,.67) -- (\x,.30);
				\node[ell cost,\style] at (\x,.92) {$\value$};
			}
			\foreach \x/\value/\style in {0/2/best,1/3/normal,2/4/normal,3/3/normal,4/2/best,5/3/normal,6/4/normal} {
				\draw[r insertion] (\x,-.67) -- (\x,-.30);
				\node[r cost,\style] at (\x,-.92) {$\value$};
			}
		\end{scope}

		\begin{scope}[xshift=7.4cm,yshift=-3.5cm]
			\node[font=\small\bfseries] at (3,1.55) {inside $V_j$ if $j\in R$};
			\foreach \x/\name in {.5/D,1.5/A,2.5/E,3.5/B,4.5/F,5.5/C}
				\node[supercandidate] at (\x,0) {$\name$};
			\node[ell cost] at (-.65,.92) {$\ell_e$};
			\node[r cost] at (-.65,-.92) {$r_e$};
			\foreach \x/\value/\style in {0/5/normal,1/4/normal,2/3/normal,3/2/normal,4/1/best,5/2/normal,6/1/best} {
				\draw[ell insertion] (\x,.67) -- (\x,.30);
				\node[ell cost,\style] at (\x,.92) {$\value$};
			}
			\foreach \x/\value/\style in {0/2/normal,1/1/best,2/2/normal,3/3/normal,4/4/normal,5/5/normal,6/4/normal} {
				\draw[r insertion] (\x,-.67) -- (\x,-.30);
				\node[r cost,\style] at (\x,-.92) {$\value$};
			}
		\end{scope}
	\end{tikzpicture}
	\medskip
	\caption{The cost of inserting an edge candidate into a vertex gadget. Each displayed number is the cost against the six supercandidates in the pictured gadget, in units of $M$. Circled values are minimal within their panel.}
	\label{fig:edge-insertion-costs}
\end{figure}
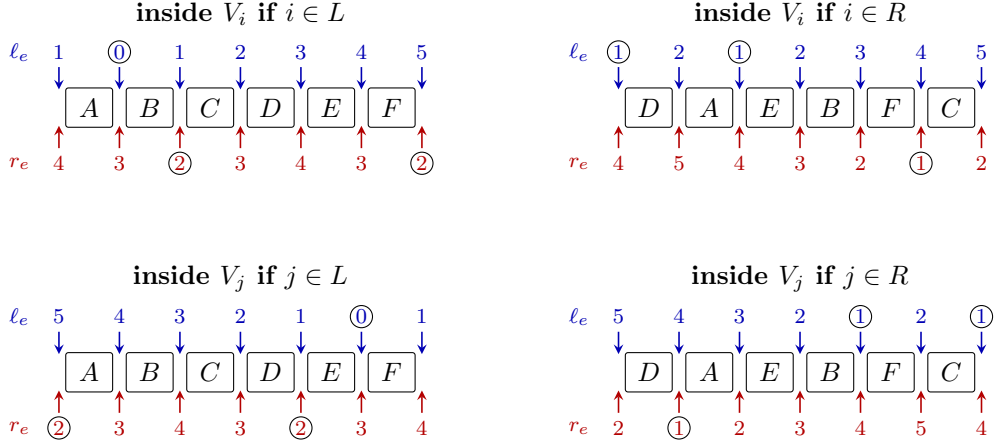

Thus, to find the lowest-cost position for an edge candidate to be placed, it suffices to pick the position inside the gadgets $V_i \cup V_j$ that minimizes the cost \emph{against candidates in the gadget where the candidate is placed}. There are seven possible insertion positions inside $V_i$ (just before $V_i$, the five positions in its interior, just after $V_i$) and seven possible insertion positions inside $V_j$. For each of these positions, we can compute the cost that the edge candidate incurs against the candidates in the vertex gadgets. These costs are shown in \Cref{fig:edge-insertion-costs}.

We see that for a candidate $\ell_e$, the lowest-cost position has cost $0$ in case one of $e$'s endpoints is placed in $L$; if neither endpoint is in $L$, then the lowest-cost position has cost $1M$. Similarly, for candidate $r_e$, the lowest-cost position has cost $1M$ in case one of $e$'s endpoints is placed in $R$; if neither endpoint is in $R$, then the lowest-cost position has cost $2M$.

Thus, placing the two candidates $\ell_e$ and $r_e$ belonging to edge $e$ in their lowest-cost positions incurs the following cost against all vertex candidates:
\[
\underbrace{(3d+2)M}_{\text{fixed cost of $\ell_e$}}
+
\underbrace{(3d-1)M + 1M}_{\text{fixed cost of $r_e$}}
+
\begin{cases}
	0 & \text{if edge $e$ is cut,} \\
	1M & \text{if it is not cut}.
\end{cases}
\]
A shorter way of writing this is
\[
(6d+3)M
-
\begin{cases}
	1M & \text{if edge $e$ is cut,} \\
	0 & \text{if it is not cut}.
\end{cases}
\]

Let $c$ be the number of edges crossing the cut $L\cup R$. By summing the above quantity over all edges, we see that placing every edge candidate in its lowest-cost position incurs the following cost between edge and vertex candidates:
\[
	\sum_{\mathclap{\{i,j\}\in\mathcal E,\ i<j}}\: (6\,(j-i)+3)M
	- cM.
\]
To recap the logic thus far: we were given an optimal Kemeny ranking $\succ$ with cost at most $b$. We saw that we can read off a cut $L \cup R$ from that ranking, and we write $c$ for the number of edges crossing the cut. We now wish to conclude that $c \ge k$.
Suppose instead that $k > c$, and so $k - 1 \ge c$. Then the cost of the ranking $\succ$ is at least
\[
3nM^2
\quad
+ 
\quad
\sum_{\mathclap{\{i,j\}\in \mathcal{E},\ i<j}} \: \bigl(6\,(j-i)+3\bigr)M - cM 
\quad + \quad
\underbrace{0}_{\clap{\parbox[c]{4cm}{\centering \scriptsize lower bound on cost \\ between edge candidates}}}
\]
which is strictly more than $b$ because $-cM \ge -(k-1)M = -kM + M > -kM + n^4$, contradicting that the cost of $\succ$ is at most $b$. Hence indeed $c \ge k$, and we are done.

\section{Three Voters Suffice}

It remains to realize the tournament used above by three voters. 
Our vertex gadget tournament (\Cref{fig:vertex-gadget-tournament}) was specifically designed to be inducible with three voters:
\[
AEFBCD, \qquad BDFACE, \qquad CDEABF.
\]
They induce exactly the 15 arcs of the tournament, each with a 2-to-1 majority.
This allows us to induce the vertex candidates and their arcs correctly using three voters:
For a permutation $\sigma$ of the letters $A,\dots,F$, write $V_i[\sigma]$ for the corresponding order of the six supercandidates of $V_i$, with the $M$ constituent candidates of each supercandidate in their natural order. Also write $\overline{V_i}[\sigma]$ for the same ordering, except that each supercandidate is internally ranked in the reverse of its natural order. Start with the following three orders of the vertex candidates:
\begin{alignat*}{4}
	\succ_1 &: V_1[AEFBCD] \quad V_2[AEFBCD] &&\;\cdots\; V_n[AEFBCD],\\
	\succ_2 &: V_n[BDFACE] \quad V_{n-1}[BDFACE] &&\;\cdots\; V_1[BDFACE],\\
	\succ_3 &: \overline{V_1}[CDEABF] \quad \overline{V_2}[CDEABF] &&\;\cdots\; \overline{V_n}[CDEABF].
\end{alignat*}
For each $V_i$, these induce the correct tournament, and they induce the majority relation between vertex gadgets as $V_1\to\cdots\to V_n$ (plus the transitive closure). Note that all majority margins between supercandidates are $\pm 1$, and the same is true within a single supercandidate because voter 3 reverses the natural order inside the supercandidate.

For every edge $e=\{i,j\}$ with $i<j$, write $d=j-i$ and let $q=(i+j)/2$ be its midpoint. Insert its edge candidates $\ell_e$ and $r_e$ into the above rankings as follows:
\[
\begin{array}{c|ccc}
	&\succ_1&\succ_2&\succ_3\\
	\midrule
	\ell_e&\text{after $A_i$ in $V_i$}&\text{after $F_q$ in $V_q$}&\text{after $B_j$ in $V_j$}\\
	r_e&\text{after $B_i$ in $V_i$}&\text{after $F_q$ in $V_q$}&\text{after $D_j$ in $V_j$}
\end{array}
\]
If several edge candidates use the same gap, break ties arbitrarily. Now we need to check that these choices induce precisely the beating sets in \Cref{tbl:edge-candidate-beaten}. Let's do it.
\begin{itemize}
	\item For $u = 1, \dots, i-1$, voters 1 and 3 rank $V_u$ above $V_i$ and hence above $\ell_e$ and $r_e$. Thus $V_u \to \{\ell_e, r_e\}$.
	\item Restricted to $V_i$, we have the orderings ${\succ_1} = A_i \ell_e E_iF_iB_i r_e C_i D_i$, ${\succ_2} = \{\ell_e,r_e\}V_i$, and ${\succ_3} = V_i\{\ell_e,r_e\}$ (where the ordering of $\ell_e$ vs.\ $r_e$ in the latter two is not specified). Note that voters 2 and 3 cancel out (w.r.t. the comparisons between $\{\ell_e,r_e\}$ and $V_i$). Hence, we can read off the majority relation from voter 1, so we have $A_i \to \ell_e \to \{B_i, \dots, F_i\}$, implementing the specification from \Cref{tbl:edge-candidate-beaten} that $\ell_e$ is beaten exactly by $A_i$. Similarly, we get $\{A_i,B_i,E_i,F_i\}\to r_e \to\{C_i,D_i\}$.
	\item For the next $(\frac{d}{2} - 1)$ gadgets $V_u$ (strictly between $i$ and the midpoint $q$), voters 1 and 2 rank $\ell_e$ and $r_e$ above $V_u$, hence $\{\ell_e, r_e\} \to V_u$.
	\item Restricted to $V_q$, we have the orderings ${\succ_1} = \{\ell_e,r_e\}V_q$, ${\succ_2} = B_qD_qF_q\{\ell_e,r_e\}A_qC_qE_q$, and ${\succ_3} = V_q\{\ell_e,r_e\}$. Voters 1 and 3 cancel out, and voter 2 implies that $\{B_q,D_q,F_q\}\to\{\ell_e,r_e\}\to\{A_q,C_q,E_q\}$.
	\item For the next $(\frac{d}{2} - 1)$ gadgets $V_u$ (strictly between the midpoint $q$ and $j$), voters 2 and 3 rank $V_u$ above $\ell_e$ and $r_e$, hence $V_u\to\{\ell_e,r_e\}$.
	\item Restricted to $V_j$, we have the orderings ${\succ_1} = \{\ell_e,r_e\}V_j$, ${\succ_2} = V_j\{\ell_e,r_e\}$, and ${\succ_3} = C_jD_jr_eE_jA_jB_j\ell_e F_j$. Voters 1 and 2 cancel out, and voter 3 implies that $\{A_j,B_j,C_j,D_j,E_j\}\to\ell_e \to F_j$, and that $\{C_j,D_j\}\to r_e \to \{A_j,B_j,E_j,F_j\}$.
	\item Finally, for $u=j+1,\dots,n$, voters 1 and 3 rank $\ell_e$ and $r_e$ above $V_u$. Thus $\{\ell_e,r_e\}\to V_u$.
\end{itemize}
Note that these arguments also show that the majority margins of all of these arcs are 1.

The final relationships to induce are the arcs between edge candidates. However, our reduction did not use or specify those arcs, so the only thing we need to check is that the majority margins between edge candidates are all $\pm 1$ and never $\pm 3$, or in other words, that there is no pair of edge candidates that all three voters rank the same way. This is certainly true for the two edge candidates associated to the same edge (because $\ell_e \succ_1 r_e$ but $r_e \succ_3 \ell_e$). So assuming for contradiction that there was such a unanimously ranked pair, there would need to be different edges $e = \{i, j\}$ and $e' = \{i', j'\}$ such that all voters put $x_e \succ_t y_{e'}$ ($t = 1,2,3$), where $x$ and $y$ are labels taken from $\{\ell, r\}$. Looking at voter 1, $x_e$ is placed in $V_i$ and $y_{e'}$ is placed in $V_{i'}$. Because $x_e \succ_1 y_{e'}$ we must have $V_{i} \succeq_1 V_{i'}$, so we deduce that $i \le i'$. Similarly, in voter 3, $x_e$ is placed in $V_j$ and $y_{e'}$ is placed in $V_{j'}$ and we deduce that $j \le j'$. Note that because $e$ and $e'$ are different edges, at least one of the inequalities $i \le i'$ and $j \le j'$ is strict. Hence, writing $q = \frac{i+j}{2}$ and $q' = \frac{i'+j'}{2}$ for the midpoints, we have $q < q'$. But then voter 2 places $x_e$ in $V_q$ and $y_{e'}$ in $V_{q'}$, and voter 2 ranks vertex gadgets oppositely, i.e., $V_{q'} \succ_2 V_q$, and hence $y_{e'} \succ_2 x_e$, contradicting unanimity.

This proves that the unit-weight tournament used in the reduction is induced by three voters; the voters also determine the previously unspecified arcs between pairs of edge candidates.

\section{Implications}

We have shown that KEMENY SCORE is hard even for three voters. From this, we can immediately deduce that it is hard for every fixed odd $n \ge 3$, because we can add two voters with completely reversed rankings to a profile without changing any majority margin. Also, KEMENY SCORE is hard for every fixed even $n \ge 4$ \citep{dwork2001rank,biedl2009complexity}. Putting these results together, we now know that KEMENY SCORE is easy for $n \in \{1,2\}$ and hard for every other fixed $n \ge 3$.

In the wider social choice literature, several works have proved hardness results by reduction from the KEMENY SCORE problem with four voters, and thereby obtained hardness results for other computational problems that hold even for a fixed small number of voters. In several cases, the same reduction can now be started from the three-voter version, thereby improving the constant for other problems too.

One example is the \emph{Egalitarian Kemeny} rule, which selects a ranking such that the maximum distance to any voter is minimized. \citet{biedl2009complexity} proved that it is NP-complete to decide whether there exists a ranking whose maximum Kendall tau distance to an input ranking is at most a given number $b$ \citep[see also][]{popov2007multiple}, and that this problem is NP-hard even for $n = 4$ voters. One can now use the same pattern as their reduction to show that the problem is hard even for $n = 3$ voters. The \emph{Squared Kemeny} rule \citep{lederer2024squared} is the rule that minimizes the sum of squared Kendall tau distances, which behaves somewhat like an average while the Kemeny rule behaves like a median. \citet{lederer2024squared} prove that computing Squared Kemeny is hard for $n = 4$ voters; one can now deduce it is hard for $n = 3$ voters by adapting their proof. 

\citet[Theorem 6.9 and Remark 6.10]{erdelyi2017nearlysp} studied the problem of deciding whether a given preference profile can be made \emph{single-peaked} using at most a given number of swaps in the profile. They showed the problem is NP-hard for 8 voters via a reduction from 4-voter Kemeny. When reducing from 3-voter Kemeny instead, one obtains NP-hardness of the global swap nearly single-peakedness detection problem even for 6 voters.

\citet[Proposition 7]{faliszewski2019similar} study whether two preference profiles can be made to be similar in terms of Kendall tau distance after relabeling voters and candidates. This problem is now known to be hard even for $n = 3$ voters.

\citet{kraiczy2023weakly} study the concept of \emph{popular rankings}, which are rankings such that no other ranking is preferred by a (simple or absolute) majority of voters, where voters prefer rankings that are Kendall tau closer to them. They give several reductions connecting this problem to the Kemeny problem for three voters. Since the latter problem is now known to be NP-hard, we can deduce that unless P = NP, one cannot decide in polynomial time whether a popular ranking exists, even for $k \in \{4,5\}$ voters.

The Kemeny method is also sometimes interpreted as a \emph{voting method}, where the highest-ranked candidate in an optimal Kemeny ranking is taken to be the winning candidate. The problem of deciding whether a given candidate is a Kemeny winner is $\Theta_2^p$-complete \citep{hemaspaandra2005complexity}; by the same subdivision argument as for the score problem (see \Cref{fn:subdivision}), this problem is also $\Theta_2^p$-complete when the number of voters is fixed to $n \ge 4$ and even. Building on the work of \citet{bachmeier2019k}, \citet{lampis2022slater} showed that this problem is $\Theta_2^p$-complete even for $n = 7$ voters (and the same reduction also shows that determining winners is hard for the closely-related \emph{Slater} voting rule, see \citealp{hudry2009slater}). It seems likely that the ideas of \citet{lampis2022slater} can be applied to the reduction in this paper to get $\Theta_2^p$-completeness of the winner determination problem even for $n = 3$.

\section*{Acknowledgements}
This work was funded in part by the Agence Nationale de la Recherche as part of the France 2030 program under grant ANR-23-IACL-0008 (PR[AI]RIE-PSAI).

\bibliography{kemeny-three-voters}

\end{document}